\documentclass[pre,10pt,preprintnumbers,superscriptaddress,twocolumn]{revtex4-2}

\usepackage{amsmath}
\usepackage{hyperref}
\usepackage{graphicx}
\usepackage{times}

\hypersetup{
    colorlinks=true,       
    linkcolor=red,          
    citecolor=blue,        
    filecolor=magenta,      
    urlcolor=cyan           
}

\newcommand{\eg}{{e.g., }}
\newcommand{\ie}{{i.e., }}

\newcommand{\brho}{{\bar{\rho}}}
\newcommand{\tf}{t_\mathrm{f}}
\newcommand{\R}{\mathrm{R}}
\renewcommand{\L}{\mathrm{L}}

\let\ccc\c
\renewcommand{\c}{\mathrm{c}}
\renewcommand{\a}{\mathrm{a}}
\newcommand{\q}{\mathrm{q}}
\newcommand{\id}{\mathrm{id}}
\newcommand{\crr}{\mathrm{cr}}
\renewcommand{\Pr}{\mathrm{Pr}}
\newcommand{\erfc}{\mathrm{erfc}}

\begin{document}

\title{Uphill drift in the absence of current in single-file diffusion}
\author{Benjamin \surname{Sorkin}}
\affiliation{School of Chemistry and Center for Physics and Chemistry of Living Systems, Tel Aviv University, 69978 Tel Aviv, Israel}
\author{David S. \surname{Dean}}
\email{david.dean@u-bordeaux.fr}
\affiliation{Univ. Bordeaux, CNRS, LOMA, UMR 5798, F-33400 Talence, France}
\affiliation{Team MONC, INRIA Bordeaux Sud Ouest, CNRS UMR 5251, Bordeaux INP, Univ. Bordeaux, F-33400 Talence, France}

\begin{abstract} 
Single-file diffusion is a paradigmatic model for the transport of Brownian colloidal particles in narrow one-dimensional channels, such as those found in certain porous media, where the particles cannot cross each other. We consider a system where a different external uniform potential is present to the right and left of an origin. For example, this is the case when two channels meeting at the origin have different radii. In equilibrium, the chemical potential of the particles are equal, the density is thus lower in the region with the higher potential, and by definition there is no net current in the system. Remarkably, a single-file tracer particle initially located at the origin, with position denoted by $Y(t)$, exhibits an average {\em up-hill} drift toward the region of {\em highest} potential. This drift has the late time behavior $\langle Y(t)\rangle= C t^{1/4}$, where the prefactor $C$ depends on the initial particle arrangement. This surprising result is shown analytically by computing the first two moments of $Y(t)$ through a simple and physically-illuminating method, and also via extensive numerical simulations.
\end{abstract}

\maketitle

Single-file diffusion (SFD) is a model for the dynamics of Brownian particles in a vast range of physical systems where transport is effectively one dimensional. It describes particles with hard core repulsion that cannot bypass each other, even in the point-like limit. SFD has been extensively studied from both theoretical ~\cite{BD04,AP78,KM12,HA65,SP70,AR83,LI83,LA08,LB13,DGL85,KMS14,KOL03,PM08,KR92,DG09,BJC22} and experimental~\cite{KR92,WCL00,LKB04,LKB04,NKP13,LBE09,YYKJOCHK10,EHZCBDFJLHCPE13,LIN05} perspectives.
This model plays a pivotal role in characterizing transport in porous media~\cite{KR92} and along the cytoskeleton~\cite{NKP13}, motion of colloids~\cite{WCL00,LKB04}, drug delievery devices~\cite{YYKJOCHK10}, and even gene regulation~\cite{LBE09}.

\begin{figure}
    \centering
    \includegraphics[width=0.95\linewidth]{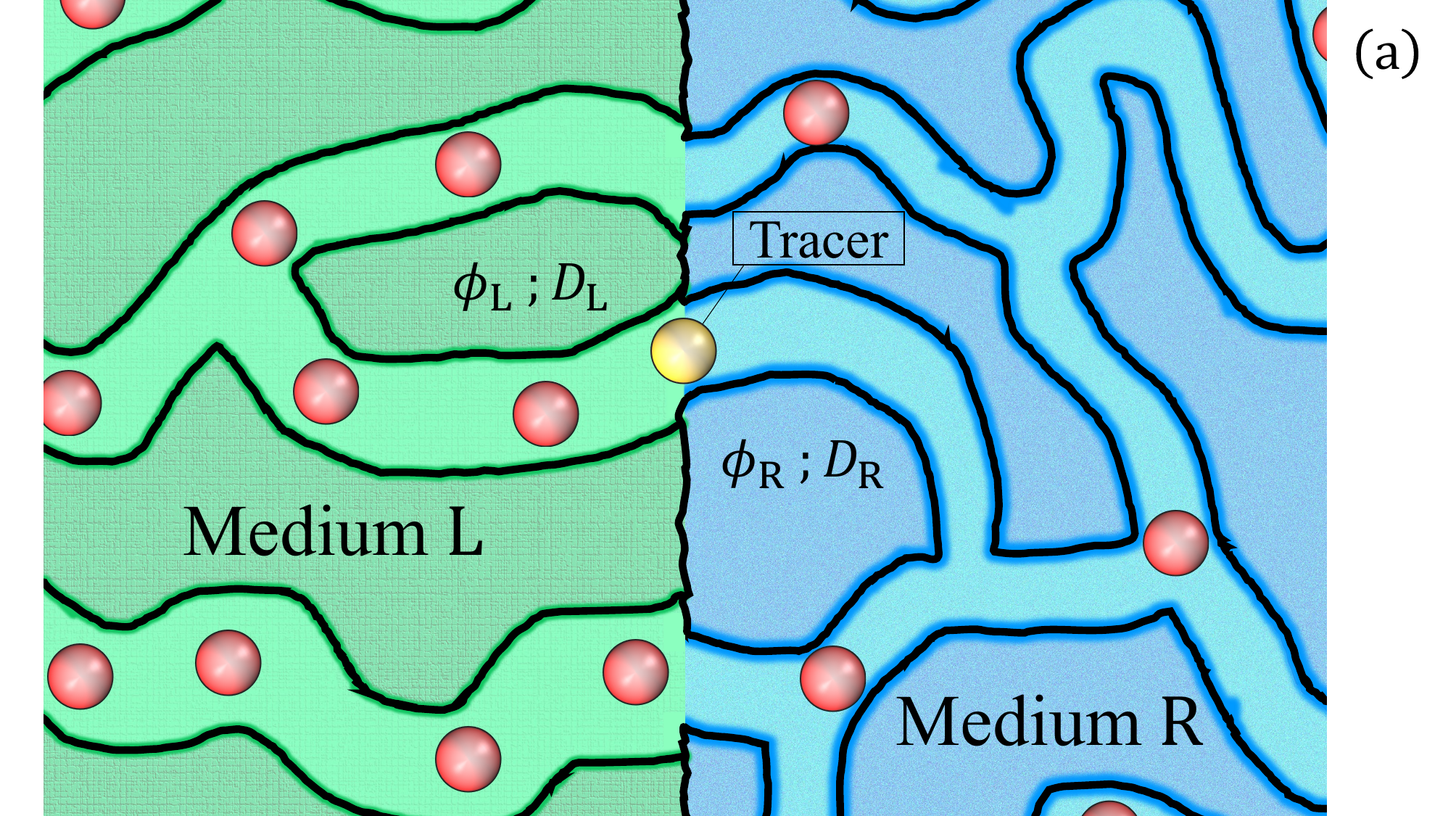}\linebreak
    \includegraphics[width=0.95\linewidth]{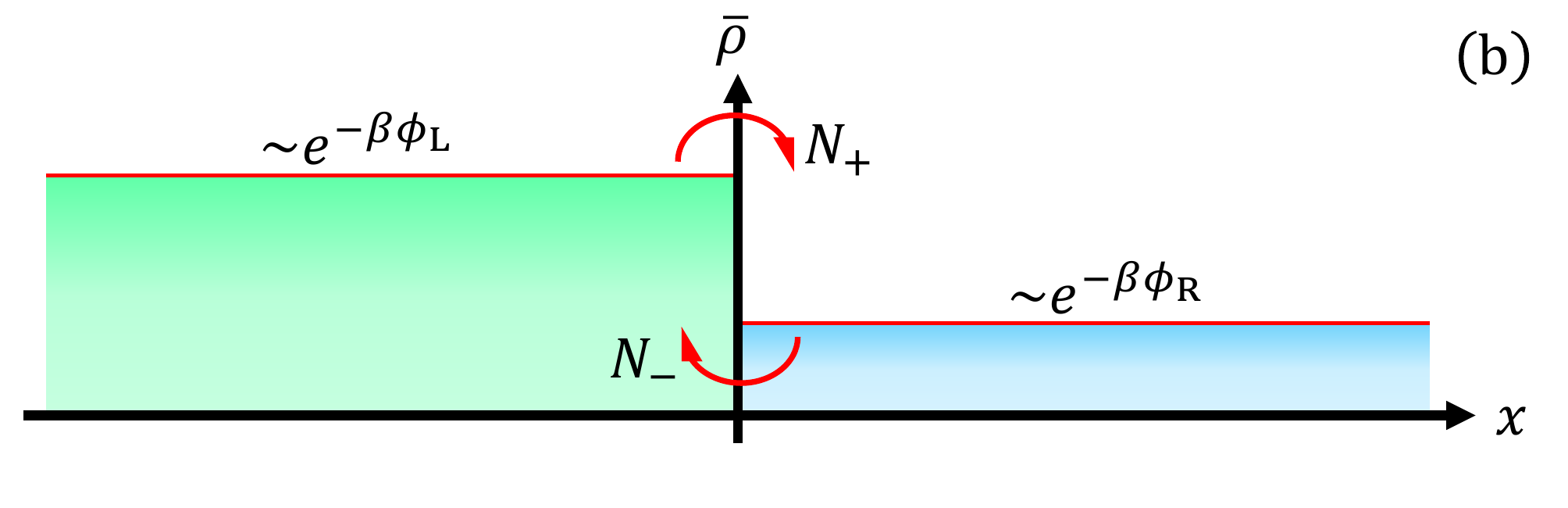}\linebreak
    \caption{(a) Illustration of an interface between two particle-laden porous media at step-like equilibrium conditions. The pores are so narrow that particles cannot overtake one another~\cite{KR92}. The pores within medium $\L$ (left) are wider than in medium $\R$ (right). (b) The average one-dimensional concentration profile $\overline\rho(x)$ across the interface. In equilibrium, the medium with the larger pores has the larger particle concentration according to $e^{-\beta(\phi_\R-\phi_\L)}=(R_\R-a)^2/(R_\L-a)^2$, where $R_{\L/\R}$ and $a$ are the pores' and the tracer's radii. The average number of particles crossings from left to right, $\langle N_+\rangle$, and from right to left, $\langle N_-\rangle$, are equal in equilibrium. }
    \label{fig:porous}
\end{figure}

From a theoretical point of view, SFD is an important model for understanding the out-of-equilibrium dynamics of interacting particle systems given its analytical tractability. For a long time it has been known that a tracer particle in SFD exhibits anomalous diffusion, having a mean-squared displacement (MSD) which grows as $t^{1/2}$~\cite{HA65,AR83,AP78}. This dramatic slowing down of dispersion is due to the dynamical caging effect by the surrounding particles. SFD also exhibits another fascinating statistical phenomena. It shows an everlasting dependence on the statistics of the initial conditions~\cite{BJC22,LB13}, as well as a dependence on how the two averages over thermal noise and initial conditions are performed~\cite{DG09,KMS14}.

SFD has been extensively studied for homogeneous environments. However, substrates like porous media and DNA strands are inherently inhomogeneous due to varying pore structures and codon sequences. This heterogeneity is exploited in applications, such as tailoring pore sizes for controlled drug release~\cite{YYKJOCHK10} and influencing ``obstacle'' protein binding rates to affect tracer sliding speed~\cite{LBE09}. Thus, investigating SFD in inhomogeneous environments holds significant interest, as it may uncover new useful phenomena.

Only a limited literature addresses SFD with space-dependent external potentials and diffusivities. The few techniques connect the tracer's displacement to the statistics of background particle crossings~\cite{KMS14,SD15,DMS23,SD23} or to the distribution of an isolated particle~\cite{P74,BS09,BS10,HSD14,CK17} for systems that are symmetric about the origin. In these cases, no new SFD-related effects have been observed other than the above, as at late times they either behave like SFD in an homogeneous environment but with a rescaled diffusion constant~\cite{KMS14,BS09,SD23} or reach the boundaries of a confining potential~\cite{BS09,BS10}. The surprising phenomenon we discover in this letter is facilitated by a single step potential arising at the interface between two differing substrates.

To illustrate, shown in Fig.~\ref{fig:porous}(a) is a sketch of an interface between two porous media, where the pore size decreases upon going from the medium on the left ($\L$) to the right ($\R$). From entropic considerations, the equilibrium concentration of colloids per unit length will be higher in medium $\L$. Namely, in the dilute limit, the number of particles within a channel is proportional to the free volume available for a single particle, $L\pi(R_{\L/\R}-a)^2$, where $L$ is the channels' length and $a$ and $R_{\R/\L}$ are the radii of the colloid and the corresponding channel. Thus, the mean one-dimensional particle densities in each bulk are $\brho_{\L/\R}\sim(R_{\L/\R}-a)^2$, which can be recast into a Boltzmann factor with the effective potential $\phi_{\L/\R}= -2\beta^{-1} \ln(R_{\L/\R}-a)$, where $\beta=1/{k_\mathrm{B}T}$ is the inverse temperature. 
Therefore, in equilibrium, the mean densities on either side are related via
\begin{equation}
\frac{\brho_\R}{\brho_\L}=e^{-\beta(\phi_\R-\phi_\L)},\label{leteq:d}
\end{equation}
as illustrated in Fig.~\ref{fig:porous}(b). A variation in the composition of the surrounding media will also lead to unequal effective potentials, for example, via differences in the van-der-Waals interactions.
The local diffusion constants, denoted by $D_{\L/\R}$, will also depend on the local environment, and in particular on the pore radii. We will consider the case of arbitrary values for $\phi_{\L/\R}$ and $D_{\L/\R}$ which are taken to be uniform in each bulk~\cite{FOOTmft}, thus generating a step-like difference in transport properties to the left and right. 

The striking result of this letter is that, for equilibrium step-like initial condition, the position of the tracer particle $Y(t)$ at late times still has an average drift. Moreover, this drift is towards the region with the {\em higher} potential, and scales as $\langle Y(t)\rangle\sim t^{1/4}$ and occurs without a macroscopic density current. We prove this result analytically and confirm it by introducing a discrete simulation method that is capable of capturing such late-time dynamics. We  also compute the MSD of the tracer and find the SFD scaling $\langle Y^2(t)\rangle_\c\sim t^{1/2}$. The numerical prefactors for both the drift and variance can be computed exactly; as in previous studies on SFD in homogeneous systems, they exhibit an everlasting dependence on initial conditions.

\textit{Theory}.--- Theoretical approaches for the treatment of free SFD (\ie no interactions other than the hard-core repulsions) include, \eg the macroscopic fluctuation theory~\cite{KMS14,BDGJL01,BDGJL05,BDGJ06} and the Bethe \textit{Ansatz} for the full joint probability density function~\cite{RKH98,PM08}.  Another, simpler approach adopted here is based on a link between SFD and two independent effusion problems~\cite{DGL85,DMS23}. This approach precisely pinpoints the physical mechanism leading to the uphill drift.

The key idea in the approach of Refs.~\cite{HA65,DGL85} is that when single-file particles cross, the hard core constraint can be imposed by relabelling the particles. This means that if all the particles are assumed to be indistinct, the system appears as if the particles do not interact with each other but only with the externally applied potentials. 
Each particle in the noninteracting system has a probability density function $p(x,t)$ which evolves according to the Fokker-Planck equation 
\begin{equation}
\frac{\partial p(x,t)}{\partial t} =\frac{\partial}{\partial x}\left\{D(x)\left[\frac{\partial p(x,t)}{\partial x}+\beta p(x,t)\frac{d \phi(x)}{dx}\right]\right\}.\label{leteq:FPE}
\end{equation}
$D(x<0)= D_\L$, $D(x>0)= D_\R$, $\phi(x<0)= \phi_\L$, and $\phi(x>0)= \phi_\R$ represents our model for diffusion of noninteracting particles in two joined channels.
Now, if the single-file tracer particle is started at $Y(t)=0$, then at time $t$ the number of particles to its left is conserved. This can be shown to give the condition~\cite{KMS14,DMS23,SD23,SM}
\begin{equation}
    \int_0^{Y(t)}dx\rho(x,t)=\int_{-\infty}^0dx[\rho(x,0)-\rho(x,t)].\label{leteq:connect}
\end{equation}
Here, $\rho(x,t)=\sum_{n=1}^N\delta(x-X_n(t))$ is the stochastic density field of the $N$ background particles, positioned at $\{X_n(t)\}$.  

We first simplify the right-hand side of Eq.~\eqref{leteq:connect}. We separate the number density field as $\rho(x,t)=\rho_\L(x,t)+\rho_\R(x,t)$ where, by keeping track of the particle identities, $\rho_\L(x,t)$ [$\rho_\R(x,t)$] corresponds to the particles that  start in the left (right) medium at $t=0$. By definition, $\rho_\L(x>0,0)=0$ [$\rho_\R(x<0,0)=0$]. Therefore, $N_+(t)$ [$N_-(t)$], the number of particle that started from the left (right) of the interface at $t=0$ and appear in the right (left) medium at time $t$, is given by $N_+(t)=\int_0^\infty dx\rho_\L(x,t)$ [$N_-(t)=\int_{-\infty}^0 dx\rho_\R(x,t)$].
Notice that $\int_0^\infty\rho_\L(x,t)+\int_{-\infty}^0\rho_\L(x,t)=\int_{-\infty}^0\rho_\L(x,0)$, with which we relate the right-hand side of Eq.~\eqref{leteq:connect} to the number of crossings~\cite{SM},
\begin{equation}
    \int_{-\infty}^0dx[\rho(x,0)-\rho(x,t)]=N_+(t)-N_-(t).\label{leteq:connectRHS}
\end{equation}

We now consider the left-hand side of Eq.~\eqref{leteq:connect} and make the assumption that $|Y(t)|$ becomes large with time. This means that we can apply the law of large numbers to the left hand-side of Eq.~\eqref{leteq:connect} and replace it with its average for a fixed big $Y_t$~\cite{DGL85,DMS23}. Since the average densities  start with their equilibrium values, the average value of $\rho(x,t)$ does not evolve in time and we can write
\begin{equation}
    \int_0^{Y(t)}dx\rho(x,t)\simeq\brho_\R Y(t)\Theta[Y(t)]+\brho_\L Y(t)\Theta[-Y(t)],\label{leteq:connectLHS}
\end{equation}
where $\Theta$ is the Heaviside step function. According to the central-limit theorem (CLT), the corrections are $\mathcal{O}[\sqrt{\brho_\R Y(t)}]$ for $Y(t)>0$ and  $\mathcal{O}[\sqrt{-\brho_\L Y(t)}]$ for $Y(t)<0$.

Inserting Eqs.~\eqref{leteq:connectRHS} and~\eqref{leteq:connectLHS} in Eq.~\eqref{leteq:connect}, we find the late-time relation between the motion of the single-file tracer and the number of crossings,
\begin{equation}
    \brho_\R Y(t)\Theta[Y(t)]+\brho_\L Y(t)\Theta[-Y(t)]=N_+(t)-N_-(t).\label{leteq:connectPRE}
\end{equation}
Equation~\eqref{leteq:connectPRE} implies that a positive displacement, $Y(t)>0$, arises from more crossing rightwards than leftwards, $N_+(t)>N_-(t)$, and vice versa. We use this to rearrange Eq.~\eqref{leteq:connectPRE} so as to express $Y(t)$ explicitly in terms of the number of crossings,
\begin{multline}
    Y(t)=\frac{[N_+(t)-N_-(t)]}{\brho_\R}\Theta[N_+(t)-N_-(t)]\\-\frac{[N_-(t)-N_+(t)]}{\brho_\L}\Theta[N_-(t)-N_+(t)].\label{leteq:connectFIN}
\end{multline}
This simple equation elucidates all the phenomena we will discuss in the following. From Eq.~\eqref{leteq:connectFIN}, we will see that  $\langle {\rm sign}[Y(t)]\rangle=0$, corresponding to zero current (the tracer is equally likely to go to the left or right) while the difference in ${\brho_\R}$ and ${\brho_\L}$ \{and the appearance of the functions  $\Theta[N_+(t)-N_-(t)]$ and $\Theta[N_-(t)-N_+(t)]$\} leads to a non-zero drift. 
In other words, while the crossings to either right [$N_+(t)>N_-(t)$] or left [$N_+(t)<N_-(t)$] occur with the same probabilities at equilibrium, the distance that the tracer moves into each bulk per crossing is bigger in the more dilute region. This point succinctly explains the physical mechanism behind the surprising non-zero tracer uphill drift without current that we report here. The rest of the analysis relies on obtaining the statistics of $Y(t)$ in terms of the two, known, independent statistics of $N_\pm(t)$~\cite{DMS23}.

In what follows, we will consider two types of initial conditions. The first is ideal-gas initial conditions, where the system is in perfect equilibrium having identical and independent uniform distribution for all the particles according to each bulk's density~\cite{FOOTid}. 
The average over the full statistics of the equilibrium initial configuration in this case will be denoted by $\langle\rangle_\id$. The second is perfect crystalline initial conditions, where the particles are set up with the equilibrium densities to the left and right of the origin but are equally spaced in a lattice, whose averages we will denote as $\langle\rangle_\crr$. 
In both cases we write the average initial densities to the left and right in the equilibrium form $\overline \rho_{\L/\R}= \overline\rho_0e^{-\beta \phi_{\L/\R}}$.

We shall compute the first two moments, $\langle Y(t)\rangle$ and $\langle Y^2(t)\rangle_\c=\langle Y^2(t)\rangle-\langle Y(t)\rangle^2$ of the single-file tracer's motion. For the ideal-gas initial conditions, we find~\cite{SM}
\begin{eqnarray}
    \langle Y(t)\rangle_{\id}&=&[e^{\beta(\phi_\R-\phi_\L)}-1]\sqrt{\frac{2}{\brho_0}\sqrt{\frac{D_1}{\pi}}}t^{1/4},\label{leteq:sfdDRIFTa}\\
    \langle Y^2(t)\rangle_{\c,\id}&=&\frac{2}{\brho_0}\sqrt{\frac{D_2}{\pi}}t^{1/2},\label{leteq:sfdMSDa}
\end{eqnarray}
where
\begin{eqnarray}
    D_1&=&\frac{D_\R D_\L}{\pi^2[e^{\beta(\phi_\R -\phi_\L)}\sqrt{D_\L}+\sqrt{D_\R}]^2},\\
    D_2&=&D_1\{[e^{2\beta(\phi_\R-\phi_\L)}+1](\pi-1)+2e^{\beta(\phi_\R-\phi_\L)}\}^2.\quad
\end{eqnarray}
On the other hand, for the crystalline initial conditions (and the quenched statistics~\cite{SM}) we find the simple relations
\begin{eqnarray}
\langle Y(t)\rangle_{\crr}&=&\langle Y(t)\rangle_{\id}/2^{1/4},\label{leteq:sfdDRIFTq}\\
\langle Y^2(t)\rangle_{\c,\crr}&=&\langle Y^2(t)\rangle_{\c,\id}/2^{1/2},\label{leteq:sfdMSDq}
\end{eqnarray}
which generically appear in SFD problems~\cite{BJC22,LB13,DG09,KMS14}.
We see that the drift in both cases is in the direction of the higher potential, thus confirming the phenomenon of up-hill drift for {\em pre-equilibriated} SFD in the model of connected pores presented here. One should note that according to Eq.~\eqref{leteq:connectFIN}, in both ideal gas and crystalline cases, $\langle {\rm sign}[Y(t)]\rangle=0$, that is to say the tracer is equally likely to move to the left or right. The effective drift seen is solely due to the lesser crowding in the region of higher potential and thus excursions of the tracer into this region typically go further. This crowding was shown to play a crucial role in the gene expression mechanism~\cite{LBE09}.

Before proceeding to the simulation, we briefly examine the case of a single isolated particle placed at the origin. The Fokker-Planck equation Eq.~\eqref{leteq:FPE} can be solved~\cite{SM,Farago20} and the first two moments computed,
\begin{align}
    \langle X(t)\rangle&\,=\,2\frac{D_\R/D_\L-e^{\beta(\phi_\R-\phi_\L)}}{(D_\R/D_\L)^{1/2}+e^{\beta(\phi_\R-\phi_\L)}}\sqrt{\frac{D_\L t}{\pi}},\label{leteq:isoDRIFT}\\
    \langle X^2(t)\rangle_{\c}\,&=\,2\left\{\frac{(D_\R/D_\L)^{3/2}+e^{\beta(\phi_\R-\phi_\L)}}{(D_\R/D_\L)^{1/2}+e^{\beta(\phi_\R-\phi_\L)}}\right.\nonumber\\&\left.-\frac{2}{\pi}\left[\frac{D_\R/D_\L-e^{\beta(\phi_\R-\phi_\L)}}{(D_\R/D_\L)^{1/2}+e^{\beta(\phi_\R-\phi_\L)}}\right]^2\right\}D_\L t.\label{leteq:isoMSD}
\end{align}
Equation~\eqref{leteq:isoDRIFT} shows that a free particle in an infinite system can also drift to the right if $D_\R/D_\L - e^{\beta(\phi_\R-\phi_\L)}>0$, that is, if $D_\R$ is sufficiently large. However, a single-file tracer will always drift toward the region of higher potential at late times, regardless of the values of the diffusivities (assuming both diffusivities are non-zero).

\textit{Simulation.}---Given the rather surprising nature of our analytical predictions, we have performed numerical simulations of both the ideal-gas and crystalline initial conditions. First, note that the relative error due to the CLT decays rather slowly (as $t^{-1/4}$), implying that the finite-time effects are significant~\cite{HSD14}. 
In addition, the discontinuities in $D(x)$ and $\phi(x)$ renders the numerical simulation rather subtle~\cite{Farago20}. This pair of challenges is particularly tricky, since we need both an extensive numerical simulations to attain the late-time regime as well as a reliable short-ranged smoothing of $\phi(x)$.

To address these issues, we devised an alternative method based on a discrete random walk model that converges to Eq.~\eqref{leteq:FPE} for small lattice spacing $\epsilon$. (See details in the Supplementary Material~\cite{SM}. There we confirm that the simulation accurately reproduces the isolated particle and particle-crossing statistics.) It has two key advantages. As a lattice model, the discontinuity is just a finite change in $\phi(x)$ and $D(x)$ over a small $\epsilon$. Second, due to the self-similarity of Eq.~\eqref{leteq:FPE}, the spacing can be increased as $\epsilon\sim t^{1/2}_\mathrm{f}$ for an arbitrarily long run time $t_\mathrm{f}$ without sacrificing accuracy. Upon simulating many isolated random walkers, the SFD constraint is then simply imposed by sorting their positions~\cite{DGL85,DMS23}. To avoid finite sized effects in SFD, the number of background particles must be increased as $t^{1/2}_\mathrm{f}$, which is thus the only added cost of a longer simulation. To our knowledge, the simulation used here has not been implemented before. It  proves to be considerably faster than the underdamped method of Ref.~\cite{Farago20}.

\begin{figure}
    \centering
    \includegraphics[width=0.99\linewidth]{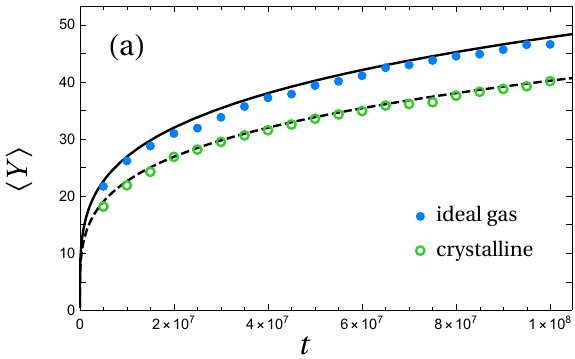}
    \includegraphics[width=0.99\linewidth]{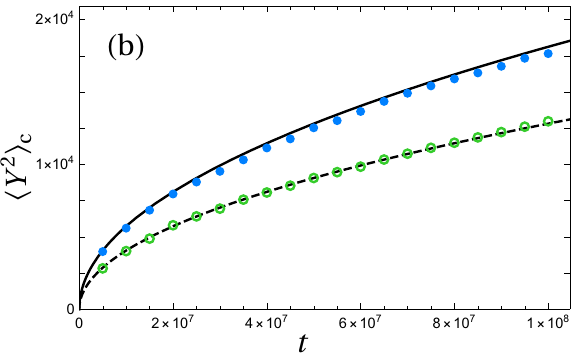}
    \caption{(a) Drift $\langle Y\rangle$ and (b) mean-squared displacement $\langle Y^2\rangle_\c$ for the single-file tracer in the longest simulation, with ideal-gas (`$\id$', blue circles) and crystalline (`$\crr$', empty green circles) initial conditions. The simulation results are depicted by the labels indicated in panel (a), while the immediately adjacent lines are the theoretical predictions [Eqs.~\eqref{leteq:sfdDRIFTa}, \eqref{leteq:sfdDRIFTq}, \eqref{leteq:sfdMSDa} and~\eqref{leteq:sfdMSDq}]. Parameter values: $D_\L=1$, $\beta\phi_\L=0$, $\brho_\L=1.59$, and $D_\R=2$, $\beta\phi_\R=1$, and $\rho_\R=0.584$. Each data point is obtained from $4\cdot10^4$ samples.}
    \label{fig:sfd}
\end{figure}

Shown in Fig.~\ref{fig:sfd} is the late-time drift and MSD for a single-file tracer for both the ideal-gas and crystalline initial conditions with $D_\R/D_\L=2$ and $\beta(\phi_\R-\phi_\L)=1$. The results of the simulation are shown by points, while the analytical late-time predictions, Eqs.~\eqref{leteq:sfdDRIFTa}, \eqref{leteq:sfdDRIFTq}, \eqref{leteq:sfdMSDa} and~\eqref{leteq:sfdMSDq}, are shown with lines. We see a good agreement with the analytical results in both cases. Most importantly, Fig.~\ref{fig:sfd}(a) unequivocally shows the existence of a mean uphill drift. 

\begin{figure}
    \centering
    \includegraphics[width=0.99\linewidth]{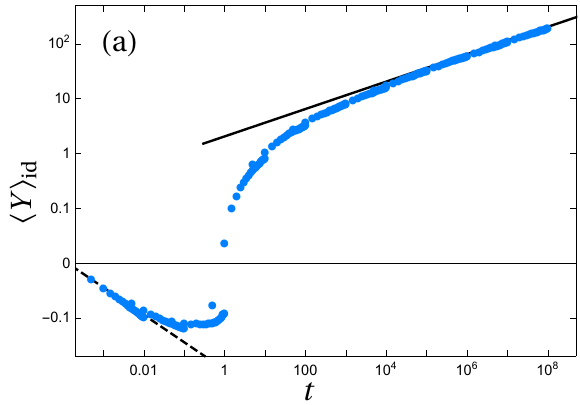}
    \includegraphics[width=0.99\linewidth]{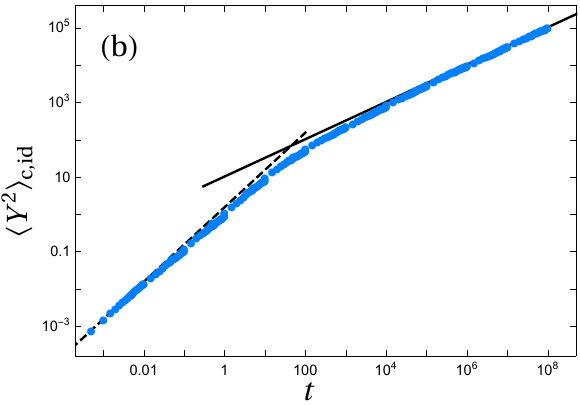}
    \caption{(a) Drift $\langle Y\rangle_{\id}$ and (b) mean-squared displacement $\langle Y^2\rangle_{\c,\id}$ for single-file tracer with ideal-gas initial conditions across many timescales. The transition from the independent normal diffusion to SFD is clearly seen in the drift turning positive after starting with a negative value. The blue points are given by the simulations, while the black lines are the theoretical predictions for the asymptotic single-file motion [Eq.~\eqref{leteq:sfdDRIFTa} and~\eqref{leteq:sfdMSDa}] and the black dashed lines are the theoretical predictions for the diffusion of an isolated particle [Eq.~\eqref{leteq:isoDRIFT} and~\eqref{leteq:isoMSD}]. Parameter values: $D_\L=1$, $\beta\phi_\L=0$, $\brho_\L=2.515$, and $D_\R=3$, $\beta\phi_\R=3$, and $\rho_\R=0.125$. Each data point is obtained from $2\cdot10^4$ samples.}
    \label{fig:sfdTRANS}
\end{figure}

In Fig.~\ref{fig:sfdTRANS} we show the drift and MSD from simulations of ideal-gas initial conditions with $D_\R/D_\L=3$ and $\beta(\phi_\R-\phi_\L)=3$, over a large range of timescales. At late times, it also demonstrates the convergence to the asymptotic analytical predictions  [Eqs.~\eqref{leteq:sfdDRIFTa} and~\eqref{leteq:sfdMSDa}]. Note that our choices for $D_\R/D_\L$ and $\beta(\phi_\R-\phi_\L)$ are a ``worse-case scenario'' for the uphill (rightward) single-file drift, as an isolated tracer starting at the origin would move downhill (or leftward) according to Eq.~\eqref{leteq:isoDRIFT}. Strikingly, indeed the short-time (free-particle-like) drift is negative as shown in Fig.~\ref{fig:sfdTRANS}(a) along with the theoretical prediction [Eq.~\eqref{leteq:isoDRIFT}], the two being in excellent agreement. Reassuringly, Fig.~\ref{fig:sfdTRANS}(b) also shows a short-time normal-diffusive regime [which agrees perfectly with Eq.~\eqref{leteq:isoMSD}, the free-particle MSD]. Thus, the tracer behaves as if it is isolated at short times, and has a downhill bias for these ``worse-case scenario'' parameters. It is only later that the single-file nature of the problem becomes dominant, at which point the uphill bias kicks in.

\textit{Conclusions.}---We showed that a single-file tracer at the junction between two channels having different potentials and diffusion constants will exhibit an effective late-time drift towards the channel of higher potential. This drift occurs without any associated  density current as the probability that the particle moves in either direction is $1/2$. This effect has been shown from both an illuminating theoretical construction as well as a late-time simulation which is particularly useful for SFD on inhomogeneous substrates. Thus, we achieve a situation where, on average, a desired tracer penetrates deeper into the higher-energy bulk without particles accumulating in that medium.

In previous studies spontaneous local drift without flux has been seen for individual Brownian particles with a spatially-varying diffusion coefficient, where the so-called spurious drift acts as a real force~\cite{LBLO01,VHBWB10,VW16}. 
A phenomenon also referred to as ``uphill diffusion'' can occur in interacting particle systems when there is a net current towards the region of higher density~\cite{CGV23}. 
The effect seen here is the opposite\,---\,motion of particles towards dilute regions (namely, ``uphill'' here refers to the hill given by the potential). We also note that for SFD in a periodic potential but driven by a constant applied force, a constant drift against the direction of the applied force has been observed in simulations~\cite{RLM19,SBK20}. However, the drift reported in this letter is a purely equilibrium one with a very different temporal dependence. 

There are potentially several situations where uphill drift may play a central role, for instance protein crowding may lead to sliding towards target regions on the DNA~\cite{LBE09}. In our treatment, we ignored the finite size of the particles, hydrodynamic effects~\cite{KOL03,LKB04}, and the finite-time corrections~\cite{SD15}, which are interesting questions for further theoretical study. Finally it should be feasible to study such systems experimentally, for instance at the junction of two microfluidic channels.

The authors acknowledge useful discussions with Haim Diamant, Eli Barkai, and Tridib Sadhu. 
D.S.D. acknowledges support  from the grant No. ANR-23-CE30-0020-01 EDIPS, and by the European Union through the European Research Council by the EMet-Brown (ERC-CoG-101039103) grant.

\renewcommand{\theequation}{S\arabic{equation}}
\setcounter{equation}{0}
\renewcommand{\thefigure}{S\arabic{figure}}
\setcounter{figure}{0}
\renewcommand{\thetable}{S\arabic{table}}
\setcounter{table}{0}
\renewcommand{\thesection}{S\Roman{section}}
\setcounter{section}{0}

\begin{widetext}
\newpage

\section*{Supplementary Material: Uphill drift in the absence of current in single-file diffusion}

In this Supplementary Material, we provide additional  technical details for the analytical results  in the main text and explain  the simulation method adopted for single-file (SF) diffusing  particles on the two sided system studied here.

\section{Analytical calculations}\label{sec:deriv}

\subsection{Single-file diffusion as an effusion problem}

We follow the standard method to relate the SF motion of a tracer, $Y(t)$, to the number of particle crossings through its initial position, $Y(0)=0$~\cite{DMS23}. First, since particles are nonpassing, the number of particles to the left of the tracer at $t=0$ and $t$ must be the same,
\begin{equation}
    \int_{-\infty}^{Y(t)}dx\rho(x,t)=\int_{-\infty}^0dx\rho(x,0),
\end{equation}
where $\rho(x,t)$ is the (stochastic) number density field. This can then be rewritten as
\begin{equation}
    \int_0^{Y(t)}dx\rho(x,t)=\int_{-\infty}^0dx\rho(x,0)-\int_{-\infty}^0dx\rho(x,t).\label{eq:connect}
\end{equation}
Equation~\eqref{eq:connect} is Eq.~(3) of the main text.

In the main text, we have defined $N_+(t)$ [$N_-(t)$]\,---\,the number of particle that started from the left (right) of the interface at $t=0$ and appear in the right (left) medium at time $t$. Separating the density field as $\rho(x,t)=\rho_\L(x,t)+\rho_\R(x,t)$, where $\rho_\L(x,t)$ [$\rho_\R(x,t)$] is the density of the particles that started from the left (right), we express the number of crossings as
\begin{equation}
    N_+(t)=\int_0^\infty dx\rho_\L(x,t),\qquad N_-(t)=\int_{-\infty}^0 dx\rho_\R(x,t).
\end{equation}
Since $\rho_\L(x>0,0)=0$ by definition, number conservation implies $\int_0^\infty\rho_\L(x,t)+\int_{-\infty}^0\rho_\L(x,t)=\int_{-\infty}^0\rho_\L(x,0)$, so $N_+(t)=\int_{-\infty}^0 dx[\rho_\L(x,0)-\rho_\L(x,t)]$. With that we relate the right-hand side of Eq.~\eqref{eq:connect} to the number of crossings,
\begin{equation}
    N_+(t)-N_-(t)=\int_{-\infty}^0dx[\rho_\L(x,0)-\rho_\L(x,t)-\rho_\R(x,t)]=\int_{-\infty}^0dx[\rho(x,0)-\rho(x,t)].\label{eq:connectRHS}
\end{equation}
Equation~\eqref{eq:connectRHS} is Eq.~\eqref{leteq:connectRHS} of the main text.

We proceed to the left-hand side of Eq.~\eqref{eq:connect}. We are dealing with average densities that correspond to the equilibrium average densities, and thus do not evolve with time. Denote $\overline{\rho(x>0,0)}=\brho_\R$ and $\overline{\rho(x<0,0)}=\brho_\L$ where, as we will define in Sec.~\ref{sec:cross} below, $\overline{\cdots}$ denotes the average over initial conditions. We make the assumption that $Y(t)$ increases with time in such a way that $\lim_{t\to\infty}|Y(t)|=\infty$. As such, after sufficiently long time, if the particle goes to the right (left), $Y(t)>0$ [$Y(t)<0$], by the the central limit theorem we find
\begin{equation}
    \int_0^{Y(t)}dx\rho(x,t)\simeq\brho_\R Y(t)\Theta[Y(t)]+\brho_\L Y(t)\Theta[-Y(t)],\label{eq:connectLHS}
\end{equation}
where $\Theta$ is the Heaviside step function. The corrections to the right-hand side are $\mathcal{O}[\sqrt{\brho_\R Y(t)}]$ and $\mathcal{O}[\sqrt{-\brho_\L Y(t)}]$ . Eq.~\eqref{eq:connectLHS} is Eq.~\eqref{leteq:connectLHS} of the main text. 

Inserting Eqs.~\eqref{eq:connectRHS} and~\eqref{eq:connectLHS} in Eq.~\eqref{eq:connect}, we find Eq.~\eqref{leteq:connectPRE} of the main text,
\begin{equation}
    \brho_\R Y(t)\Theta[Y(t)]+\brho_\L Y(t)\Theta[-Y(t)]=N_+(t)-N_-(t).\label{eq:connectPRE}
\end{equation}
which can be inverted to become an explicit relation for $Y(t)$  [Eq.~\eqref{leteq:connectFIN} of the main text],
\begin{equation}
    Y(t)=\frac{[N_+(t)-N_-(t)]}{\brho_\R}\Theta[N_+(t)-N_-(t)]-\frac{[N_-(t)-N_+(t)]}{\brho_\L}\Theta[N_-(t)-N_+(t)].\label{eq:connectFIN}
\end{equation}
Equation~\eqref{eq:connectFIN} is useful  as it allows us to determine the statistics of $Y(t)$ in terms of the statistics of the two independent effusion problems for $N_\pm(t)$~\cite{DMS23}. 

\subsection{Annealed versus quenched averages}\label{sec:cross}

We recall that SF systems possess  two types of averages~\cite{DG09}. First is the average over thermal noise which acts on the individual particles during the dynamical evolution from the starting configuration, $\langle\cdots\rangle$, and the second is an average over the particle's initial configuration, $\overline{\cdots}$. While this only produces a single first cumulant for a function $Q$,
\begin{equation}
    \langle Q(t)\rangle_\a=\langle Q(t)\rangle_\q\equiv\overline{\langle Q(t)\rangle},
\end{equation}
there are two possible definitions for the second cumulant~\cite{DG09}\,---\,annealed
\begin{equation}
    \langle Q^2(t)\rangle_{\c,\a}\equiv\overline{\langle Q^2(t)\rangle}-[\overline{\langle Q(t)\rangle}]^2,
\end{equation}
and quenched
\begin{equation}
    \langle Q^2(t)\rangle_{\c,\q}\equiv\overline{\langle Q^2(t)\rangle}-\overline{[\langle Q(t)\rangle]^2}.
\end{equation}
If  we start from equilibrium initial conditions , the particles are uniformly distributed within each bulk with densities $\brho_\L$ and $\brho_\R=e^{-\beta(\phi_\R-\phi_\L)}\brho_\L$. The quenched average is rather difficult to obtain from simulations. But, we will see later that a quenched average applied to the ideal gas case corresponds to the annealed average for perfect crystalline (or hyperuniform~\cite{sal18}) initial conditions~\cite{KMS14,BJC22}.

\subsection{Annealed and quenched generating functions for crossing statistics}

Assume the number of particles initially to the left of the interface on $[-L,0]$ is $N_\L$, where $\brho_\L= N_\L/L$. Additionally, assume that the distribution is uniform, so one has ideal gas initial conditions. The number of particles that have crossed to the right at time $t$ is given by
\begin{equation}
    N_+(t)=\sum_{n=1}^{N_\L}\Theta[X_n(t)].\label{eq:Np}
\end{equation}
We define the  moment-generating function (MGF) dependent on the initial conditions as 
\begin{equation}
    G_{N_+}(\mu,t|\{X_n(0)\})=\left\langle\exp\left\{-\mu\sum_{n=1}^{N_\L}\Theta[X_n(t)]\right\}\right\rangle,
\end{equation}
where we have only averaged over realizations of the thermal noise, not the initial configuation $\{X_m(0)\}$. 

The annealed MGF is given by
\begin{equation}
    g_{N_+,\a}(\mu,t)=\overline{G_{N_+}(\mu,t|\{X_n(0)\})},\label{eq:ganneal}
\end{equation}
and the quenched one by
\begin{equation}
    g_{N_+,\q}(\mu,t)=\exp\{\overline{\ln [G_{N_+}(\mu,t|\{X_n(0)\}])}\}.\label{eq:gquench}
\end{equation}
It is easy to see that by taking derivatives with respect to $\mu$, Eq.~\eqref{eq:gquench} generates the quenched second moment, while Eq.~\eqref{eq:ganneal} gives the annealed second moment.

Using $e^{-\mu\Theta(x)}=1-\Theta(x)+e^{-\mu}\Theta(x)$, we find
\begin{equation}
    G_{N_+}(\mu,t|\{X_n(0)\})=\prod_{n=1}^{N_\L}[1+(e^{-\mu}-1)f(X_n(0),t)],
\end{equation}
where 
\begin{equation}
    f(X_n(0),t)=\langle\Theta[X_n(t)]\rangle\label{eq:fDEF}     
\end{equation}
depends on $X_n(0)$ only since the thermal averages over each particle trajectory is independent.

\subsection{Solutions to Fokker-Planck equation}

We denote the probability density function for an isolated particle, started at $y$, around the point  $x$ as $p_X(x,t|y,0)$ which obeys the forward Fokker-Planck equation (FPE)
\begin{gather}
    \frac{\partial p_X(x,t|y,0)}{\partial t}=\hat{H}(x)p_X(x,t|y,0)\label{eq:FPE}\\\hat{H}(x)=\frac{\partial}{\partial x}\left[D(x)\left(\beta\frac{d \phi(x)}{d x}+\frac{\partial}{\partial x}\right)\right]=\frac{\partial}{\partial x}\left[e^{-\beta\phi(x)}D(x)\frac{\partial}{\partial x}\left(e^{\beta\phi(x)}\right)\right],
\end{gather}
with the initial conditions $p_X(x,0|y,0)=\delta(x-y)$, and
\begin{eqnarray}
    \phi(x)	&=&	\begin{cases}
\phi_{\L}, & x<0,\\
\phi_{\R}, & x>0,\label{eq:pot}
\end{cases}\\
D(x)	&=&	\begin{cases}
D_{\L}, & x<0,\\
D_{\R}, & x>0.\label{eq:dif}
\end{cases}
\end{eqnarray}

Note that we can write $f(y,t)$ of Eq.~\eqref{eq:fDEF} as 
\begin{equation}
    f(y,t)=\int_0^\infty dx p_X(x,t|y,0),\label{eq:fCOMP}
\end{equation}
which is an identical function to all particles and obeys the backward FPE~\cite{OKS03},
\begin{gather}
    \frac{\partial f(y,t)}{\partial t}=\hat H^\dagger(y) f(y,t),\label{eq:FPEadj}\\
    \hat H^\dagger(y)=e^{\beta\phi(y)}\frac{\partial}{\partial y}\left(D(y)e^{-\beta\phi(y)}\frac{\partial}{\partial y}\right),
\end{gather}
with the initial conditions $f(y,0)=\Theta(y)$. We define
\begin{equation}
    f(y,t)	=	\begin{cases}
f_{\L}(y,t), & y<0,\\
f_{\R}(y,t), & y>0,
\end{cases}
\end{equation}
and Laplace transform it as $\tilde f(y,s)=\int_0^\infty dt e^{-st}f(y,t)$. This yields the equations
\begin{eqnarray}
    s\tilde f_\L(y,s)&=&D_\L\frac{\partial^2 \tilde f_\L(y,s)}{\partial y^2},\\
    s\tilde f_\R(y,s)-1&=&D_\R\frac{\partial^2 \tilde f_\R(y,s)}{\partial y^2},
\end{eqnarray}
while ensuring continuity: $\tilde f_\L(0^{-},s)=\tilde f_\R(0^{+},s)$ and $D_\L e^{-\beta\phi_\L}(\partial\tilde f_\L/\partial y)|_{0^{-},s}=D_\R e^{-\beta\phi_\R}(\partial\tilde f_\R/\partial y)|_{0^{+},s}$. Upon solving these equations and inverting the Laplace transform, we find
\begin{equation}
    f_\L(y,t)\equiv F\left(\frac{y}{\sqrt{t}}\right)= \frac{\sqrt{D_\R/D_\L}}{e^{\beta(\phi_\R-\phi_\L)}+\sqrt{D_\R/D_\L}}\erfc\left(\frac{-y}{\sqrt{4D_\L t}}\right),\label{eq:FuL}
\end{equation}
where $\erfc (u)=(2/\pi^{1/2})\int_u^\infty\exp[-t^2]dt$ is the complementary error function. 

\subsection{Tracer statistics for ideal-gas initial conditions}\label{sec:annealed}

This is the case of an ideal-gas initial condition considered in the main text. Each particle is independently distributed in space with a probability density function given by $p_X(x,t)=(\brho_\L/N)\Theta(-x)+(\brho_\R/N)\Theta(x)$, where $N=N_\L+N_\R$. 

We take the average over initial conditions, and use the fact that the particles are identically and independently distributed,
\begin{equation}
    g_{N_+,\id}(\mu,t)=\left[1+(e^{-\mu}-1)\overline{f(X(0),t)}\right]^{N_\L},
\end{equation}
where each particles that has started from the left has the uniform conditional distribution, $\brho_\L/N_\L$,
\begin{equation}
    \overline{f(X(0),t)}= \frac{\brho_\L}{N_\L}\int_{-\infty}^0dyf(y,t).
\end{equation}
Upon changing variables $y\to ut^{1/2}$,
\begin{equation}
    \overline{f(X(0),t)}=\frac{\brho_\L\sqrt{t}}{N_\L}\int_{-\infty}^0duF(u).\label{eq:fRES}
\end{equation}
Thus,
\begin{equation}
    g_{N_+,\id}(\mu,t)=\left[1+(e^{-\mu}-1) \frac{\brho_\L\sqrt{t}}{N_\L}\int_{-\infty}^0duF(u)\right]^{N_\L}\xrightarrow[]{N_\L\to\infty}\exp\left[\brho_\L\sqrt{t}(e^{-\mu}-1)\int_{-\infty}^0duF(u)\right].\label{eq:gapPOIS}
\end{equation}

Now, when  $t$ becomes large, only small $\mu$ contributes. As such, we expand
\begin{equation}
    g_{N_+,\id}(\mu,t)=\exp\left[-\mu\brho_\L\sqrt{t}\int_{-\infty}^0duF(u)+\frac{\mu^2}{2}\brho_\L\sqrt{t}\int_{-\infty}^0duF(u)+\mathcal{O}(\mu^3)\right].
\end{equation}
This is exactly the MGF of a Gaussian distribution and it is in fact the manifestation of the central limit theorem in our computation. The probability distribution of $N_+$, $p_{N_+,\id}(n,t)$, is thus normal and given by
\begin{equation}
    p_{N_+,\id}(n,t)=\frac{1}{\sqrt{2\pi\sigma_{\id+}^2t^{1/2}}}\exp\left[-\frac{(n-\alpha_{\id+}t^{1/2})^2}{2\sigma_{\id+}^2t^{1/2}}\right],
\end{equation}
with
\begin{equation}
    \alpha_{\id+}=\sigma_{\id+}^2=\brho_\L\int_{-\infty}^0duF(u).\label{eq:a+}
\end{equation}

Notice that Eq.~\eqref{eq:gapPOIS} is the MGF of a Poisson distribution, which required no assumption regarding the dynamics (as was found generically in Ref.~\cite{DMS23}). Indeed a Poisson distribution with a large mean can be approximate by a Gaussian. Notice that upon inverting $\L\leftrightarrow\R$, we obtain the equilibrium  condition $\alpha_{\a+}=\alpha_{\a-}$ so long as the initial average densities satisfy the equilibrium condition, $\brho_\R=\brho_\L e^{-\beta(\phi_\R-\phi_\L)}$. In our case, denote $\brho_0=\brho_\L e^{\beta\phi_\L}=\brho_\R e^{\beta\phi_\R}$, and use Eq.~\eqref{eq:FuL} in Eq.~\eqref{eq:a+} to find
\begin{equation}
    \langle N_\pm(t)\rangle_{\id}=\langle N^2_\pm(t)\rangle_{\c,\id}=\frac{2\brho_0}{e^{\beta\phi_\R}/\sqrt{D_\R}+e^{\beta\phi_\L}/\sqrt{D_\L}}\sqrt{\frac{t}{\pi}}.\label{eq:NavA}
\end{equation}

\subsection{Crystalline initial configuration statistics}\label{sec:hyperuniform}

On the left of the origin  (the same argument applies to the right) we assume that the particles are initially placed on a grid, $X_n(0)=-(U_0+n)/\brho_\L$. Here, $U_0\in[0,1)$ is some uniform-distributed offset applied to all particles, ensuring a truly uniform particle density $\brho_\L$ (rather than a Dirac comb).
This crystalline arrangement implies that the initial configuration is not the equilibrium one, as the particles are strongly correlated with one another~\cite{sal18}. Nevertheless, the average density is equal to the equilibrium one. At late times the choice of $U_0$ will be unimportant, too, as we shall soon see.
The annealed MGF for these initial conditions is given by
\begin{equation}
    g_{N_+,\crr}(\mu,t)=\prod_{n=1}^{N_\L}\left[1+(e^{-\mu}-1)f\left(-\frac{U_0+n}{\brho_\L},t\right)\right].
\end{equation}
Rearranging it as
\begin{equation}
    g_{N_+,\crr}(\mu,t)=\exp\left\{\sum_{n=1}^{N_\L}\ln\left[1+(e^{-\mu}-1)F\left(-\frac{U_0+n}{\brho_\L \sqrt{t}}\right)\right]\right\}
\end{equation}
and taking the limit of large $t$, such that $u=-n/(\brho_\L\sqrt t)$ becomes a continuous variable [with $U_0/(\brho_\L\sqrt t)\to0$ as well], we find
\begin{equation}
    g_{N_+,\crr}(\mu,t)=\exp\left\{\brho_\L\sqrt t\int_{-\infty}^0 du\ln\left[1+(e^{-\mu}-1)F(-u)\right]\right\}.\label{eq:gcrys}
\end{equation}

Observe that Eq.~\eqref{eq:gcrys} is exactly the MGF for the quenched statistics given in Eq.~\eqref{eq:gquench} with ideal-gas initial statistics. Thus the two types of statistics indeed coincide, and give rise to the same behavior for the SF tracer. Henceforth, we will only refer to the annealed average over crystalline initial conditions, but the same results  also apply  to the quenched statistics of the idea gas initial conditions.

For large $t$ we can expand for small $\mu$ and inverse Laplace transform to find the distribution of number of crossings,
\begin{equation}
     p_{N_+,\crr}(n,t)=\frac{1}{\sqrt{2\pi\sigma_{\crr+}^2t^{1/2}}}\exp\left[-\frac{(n-\alpha_{\crr+}t^{1/2})^2}{2\sigma_{\crr+}^2t^{1/2}}\right],\label{p_+}
\end{equation}
where
\begin{eqnarray}
    \alpha_{\crr+}&=&\brho_\L\int_{-\infty}^0duF(u),\label{eq:q1+}\\
    \sigma^2_{\crr+}&=&\brho_\L\int_{-\infty}^0du[F(u)-F^2(u)].\label{eq:q2+}
\end{eqnarray}
Inserting Eq.~\eqref{eq:FuL} in Eqs.~\eqref{eq:q1+} and~\eqref{eq:q2+} and using the $\L\leftrightarrow\R$ permutation, we find the following two moments for the number of crossings:
\begin{equation}
    \langle N_\pm(t)\rangle_{\crr}=\frac{2\brho_0}{e^{\beta\phi_\R}/\sqrt{D_\R}+e^{\beta\phi_\L}/\sqrt{D_\L}}\sqrt{\frac{t}{\pi}},\label{eq:NavQ}
\end{equation}
and
\begin{eqnarray}
    \langle N^2_+(t)\rangle_{\c,\crr}&=&\frac{2\brho_0[e^{\beta\phi_\R}/\sqrt{D_\R}+(\sqrt2-1)e^{\beta\phi_\L}/\sqrt{D_\L}]}{(e^{\beta\phi_\R}/\sqrt{D_\R}+e^{\beta\phi_\L}/\sqrt{D_\L})^2}\sqrt{\frac{t}{\pi}},\label{eq:N2avQpos}\\
    \langle N^2_-(t)\rangle_{\c,\crr}&=&\frac{2\brho_0[(\sqrt2-1)e^{\beta\phi_\R}/\sqrt{D_\R}+e^{\beta\phi_\L}/\sqrt{D_\L}]}{(e^{\beta\phi_\R}/\sqrt{D_\R}+e^{\beta\phi_\L}/\sqrt{D_\L})^2}\sqrt{\frac{t}{\pi}}.\label{eq:N2avQneg}
\end{eqnarray}
It is noteworthy  that $\langle N^2_+(t)\rangle_{\c,\crr}+\langle N^2_-(t)\rangle_{\c,\crr}=2^{-1/2}[\langle N^2_+(t)\rangle_{\c,\id}+\langle N^2_-(t)\rangle_{\c,\id}]$.

\subsection{Drift and mean-squared displacement}

Now that we are equipped with the results for both the ideal-gas (annealed statistics) and crystalline initial conditions (quenched statistics), we can compute the first (drift) and second (mean-squared displacement) cumulants of the tracer motion, $Y(t)$. To this purpose, we define $\Delta N(t)=N_+(t)-N_-(t)$ and recall $\brho_0=\brho_\L e^{\beta\phi_\L}=\brho_\R e^{\beta\phi_\R}$. Then, using $\Theta^2(u)=\Theta(u)$, $\Theta(-u)=1-\Theta(u)$, and $\Theta(-u)\Theta(u)=0$, we rewrite Eq.~\eqref{eq:connectFIN} as
\begin{eqnarray}
    Y(t)&=&\frac{e^{\beta\phi_\R}-e^{\beta\phi_\L}}{\brho_0}\Delta N(t)\Theta[\Delta N(t)]+\frac{e^{\beta\phi_\L}}{\brho_0}\Delta N(t),\label{eq:driftPREP}\\
    Y^2(t)&=&\frac{e^{2\beta\phi_\R}-e^{2\beta\phi_\L}}{\brho^2_0}\Delta N^2(t)\Theta[\Delta N(t)]+\frac{e^{2\beta\phi_\L}}{\brho^2_0}\Delta N^2(t).\label{eq:msdPREP}
\end{eqnarray}

Taking the convolution of $p_{N_+,j}$ and $p_{N_-,j}$ gives the probability distribution for the difference between the number of crossings, $\Delta N(t)=N_+(t) -N_-(t) $, with $j=\id,\crr$,
\begin{equation}
    p_{\Delta N,j}(\delta n,t)=\frac{1}{\sqrt{2\pi[\langle N^2_+(t)\rangle_{\c,j}+\langle N^2_-(t)\rangle_{\c,j}]}}\exp\left\{-\frac{\delta n^2}{2[\langle N^2_+(t)\rangle_{\c,j}+\langle N^2_-(t)\rangle_{\c,j}]}\right\},\label{eq:normDN}
\end{equation}
where the mean difference in number of crossings is zero in equilibrium, and the variance is the sum of variances since the number of crossings to the right is independent of the number to the left. Thus, integrating Eqs.~\eqref{eq:driftPREP} and~\eqref{eq:msdPREP} with respect to the $i$th distribution, we find
\begin{eqnarray}
    \langle Y(t)\rangle_{j}&=&\frac{e^{\beta\phi_\R}-e^{\beta\phi_\L}}{\brho_0}\int_0^{\infty}d(\delta n)p_{\Delta N,j}(\delta n,t)\delta n+\frac{e^{\beta\phi_\L}}{\brho_0}\int_{-\infty}^{\infty}d(\delta n)p_{\Delta N,j}(\delta n,t)\delta n\nonumber\\
    &=&\frac{e^{\beta\phi_\R}-e^{\beta\phi_\L}}{\brho_0}\sqrt{\frac{\langle N^2_+(t)\rangle_{\c,j}+\langle N^2_-(t)\rangle_{\c,j}}{2\pi}},\\
    \langle Y^2(t)\rangle_{\c,j}&=&\frac{e^{2\beta\phi_\R}-e^{2\beta\phi_\L}}{\brho^2_0}\int_0^{\infty}d(\delta n)p_{\Delta N,j}(\delta n,t)\delta n^2+\frac{e^{2\beta\phi_\L}}{\brho^2_0}\int_{-\infty}^{\infty}d(\delta n)p_{\Delta N,j}(\delta n,t)\delta n^2-\langle Y(t)\rangle_{j}^2\nonumber\\
    &=&\frac{1}{2\brho^2_0}\left[{(e^{2\beta\phi_\R}+e^{2\beta\phi_\L})\left(1-\frac{1}{\pi}\right)+\frac{2}{\pi}e^{\beta(\phi_\R+\phi_\L)}}\right][\langle N^2_+(t)\rangle_{\c,j}+\langle N^2_-(t)\rangle_{\c,j}].
\end{eqnarray}

Upon inserting the variances of the number of crossings, Eqs.~\eqref{eq:NavA}, \eqref{eq:N2avQpos} and~\eqref{eq:N2avQneg}, we find the central results of our letter\,---\,the ideal-gas (annealed) and crystalline (quenched) drift and mean-squared displacements. Namely,
\begin{eqnarray}
    \langle Y(t)\rangle_{\id}&=&(e^{\beta\phi_\R}-e^{\beta\phi_\L})\sqrt{\frac{2}{\brho_0}\sqrt{\frac{D_1t}{\pi}}},\label{eq:sfdDRIFTa}\\
    \langle Y^2(t)\rangle_{\c,\id}&=&\frac{2}{\brho_0}\sqrt{\frac{D_2t}{\pi}},\label{eq:sfdMSDa}
\end{eqnarray}
where
\begin{eqnarray}
    D_1&=&\frac{1}{\pi^2(e^{\beta\phi_\R}/\sqrt{D_\R}+e^{\beta\phi_\L}/\sqrt{D_\L})^2},\label{eq:D1def}\\
    D_2&=&D_1[(e^{2\beta\phi_\R}+e^{2\beta\phi_\L})(\pi-1)+2e^{\beta(\phi_\R+\phi_\L)}]^2,
\end{eqnarray}
and
\begin{eqnarray}
    \langle Y(t)\rangle_{\crr}&=&(e^{\beta\phi_\R}-e^{\beta\phi_\L})\sqrt{\frac{\sqrt2}{\brho_0}\sqrt{\frac{D_1t}{\pi}}},\label{eq:sfdDRIFTq}\\
    \langle Y^2(t)\rangle_{\c,\crr}&=&\frac{\sqrt2}{\brho_0}\sqrt{\frac{D_2t}{\pi}}.\label{eq:sfdMSDq}
\end{eqnarray}
These are Eqs.~\eqref{leteq:sfdDRIFTa}--\eqref{leteq:sfdMSDq} of the main text.

\subsection{Late-time probability density of tracer}
We conclude this section by finding the complete probability density function  of the tracer's displacement. We do so by computing the characteristic function of $Y(t)$ for ideal gas  ($j=\id$) or perfect crystalline ($j=\crr$) statistics,
\begin{equation}
    \psi_j(q,t)=\langle e^{-iq Y(t)}\rangle_j.
\end{equation}
Using Eqs.~\eqref{eq:connectFIN} we rewrite the characteristic function as
\begin{equation}
    \psi_j(q,t)=\left\langle \exp\left[-iq\left(\frac{\Delta N(t)}{\brho_\R}\Theta[\Delta N(t)]+\frac{\Delta N(t)}{\brho_\L}\Theta[-\Delta N(t)]\right)\right]\right\rangle_j,
\end{equation}
which, using $e^{\alpha u\Theta(u)+\beta u\Theta(-u)}=e^{\alpha u}\Theta(u)+e^{\beta u}\Theta(-u)$, becomes
\begin{equation}
    \psi_j(q,t)=\langle e^{-i(q/\brho_\R)\Delta N(t)}\Theta[\Delta N(t)]\rangle_j+\langle e^{-i(q/\brho_\L)\Delta N(t)}\Theta[-\Delta N(t)]\rangle_j.
\end{equation}
With the aid of Eq.~\eqref{eq:normDN}, we find
\begin{equation}
    \psi_j(q,t)=\frac{1}{2}\exp\left(-\frac{\Sigma_j^2q^2t^{1/2}}{2\brho_\R^2}\right)\erfc\left(\frac{i\Sigma_jqt^{1/4}}{\sqrt2\brho_\R}\right)+\frac{1}{2}\exp\left(-\frac{\Sigma_j^2q^2t^{1/2}}{2\brho_\L^2}\right)\erfc\left(-\frac{i\Sigma_j qt^{1/4}}{\sqrt2\brho_\L}\right).
\end{equation}
where $\Sigma_j^2\equiv\sigma_{j+}^2+\sigma_{j-}^2$. 

An inverse Fourier transform yields an anticipated result,
\begin{equation}
   \Pr_j[Y(t)=y]=\begin{cases}
\dfrac{1}{\sqrt{2\pi \mathcal{V}_{j\L}(t)}}\exp\left[-\dfrac{y^2}{2\mathcal{V}_{j\L}(t)}\right], & y<0,\\
\dfrac{1}{\sqrt{2\pi \mathcal{V}_{j\R}(t)}}\exp\left[-\dfrac{y^2}{2\mathcal{V}_{j\R}(t)}\right], & y>0,
\end{cases}\label{eq:normDY}
\end{equation}
where, upon inserting $\Sigma_j^2$, the bulk-dependent ``variances'' are $\mathcal{V}_{\id,\L/\R}(t)=(4e^{2\beta\phi_{\L/\R}}/\brho_0)\sqrt{\pi D_1t}$ in the ideal-gas case and $\mathcal{V}_{\crr,\L/\R}(t)=\mathcal{V}_{\id,\L/\R}(t)/\sqrt2$ in the crystalline case, and $D_1$ is given in Eq.~\eqref{eq:D1def}.

Observe how the reported effect is manifested in Eq.~\eqref{eq:normDY}. The chances to go right ($\Pr[Y(t)>0]$) or left ($\Pr[Y(t)<0]$) are both $1/2$ (due to the equilibrium condition and the resulting equal number of crossings to either side). However, the range into which the tracer penetrates [$\mathcal{V}_{j\R}(t)$ versus $\mathcal{V}_{j\L}(t)$] is larger within the bulk with the higher energy.

\section{Simulation}\label{sec:simulation}

Based upon the non-crossing interpretation of SF diffusion of Ref.~\cite{DGL85}, we propose the following procedure. The $N$ particles undergoing SF diffusion will be described by $N$ independent simulations of a normally-diffusing, isolated particle which, in each of the $N$ simulations, has started from each of the initial positions of the $N$ SF particles chosen positions. At any time when  wish to record the positions of the SF particles, we sort the array of the normally-diffusing particle across the $N$ (``phantom'') simulations, thus achieving the sought after relabeling. Using the independent simulations, we are also able to confirm the scheme for the diffusion of  non-interacting particles and as well as  verify the number crossings statistics for these particles.

At this point, we develop  a simulation method for the two-sided discontinuous system under study here. Typically, one considers potentials and diffusivities that vary smoothly in space, so one can resort to simple overdamped Langevin dynamics for the individual particles. However, the literature on simulation methods of diffusion processes across discontinuities is limited~\cite{Farago20}. 

To our knowledge, the lattice method derived below has not been implemented before in the context of SFD. For the case considered here, this  method turns out to be significantly faster than the continuous method of Ref.~\cite{Farago20}. The latter is based on the use  of an effective underdamped equation of motion where, intricately-structured potentials are be properly sampled with a sufficiently fine temporal discretization at the  scale where ballistic motion dominates. As such, the potential discontinuity is addressed by smoothing it out and properly sampling it. Likewise, our lattice method is, in theory, restricted to a spatial discretization which is much finer that the spatial variability of the potential and diffusivity. As a lattice method, however, discontinuities do not require special attention compared to smooth potentials. Combined with the self-similarity of diffusion through a single interface [see Eq.~\eqref{eq:sol_NORMAL} below], there is, in fact, no limitation on the discretization (see Sec.~\ref{sec:walkthrough}). Thus, a very coarse discretization for long-time simulations in the following method turns out to be faster by orders of magnitudes than Ref.~\cite{Farago20}.

\subsection{Derivation of the single-particle simulation}

Our method is a simulation of a particle on a lattice. When the particle is at site $n$, it attempts to move to the right with rate
\begin{equation}
W_{n\to n+1} = \lambda_{n}\frac{e^{-\beta \phi_{n+1}}}{e^{-\beta \phi_{n}} + e^{-\beta \phi_{n+1}}},\label{eq:rate_W}
\end{equation} 
while the reverse move to the left occurs with rate
\begin{equation}
W_{n+1\to n} = \lambda_{n+1}'\frac{e^{-\beta \phi_{n}}}{e^{-\beta \phi_{n}} + e^{-\beta \phi_{n+1}}}.\label{eq:rate_Wrev}
\end{equation}
In order to satisfy detailed balance, $W_{n\to n+1}/W_{n+1\to n}=e^{-\beta \phi_{n+1}}/e^{-\beta \phi_{n}}$, we require $\lambda_{n+1}'=\lambda_n$. The resulting master equation is
\begin{eqnarray}
\frac{\partial p_n(t)}{\partial t} &=& -p_n(t)\lambda_n\frac{e^{-\beta \phi_{n+1}}}{e^{-\beta \phi_{n}} + e^{-\beta \phi_{n+1}}}- p_n(t)\lambda_n'\frac{e^{-\beta \phi_{n-1}}}{e^{-\beta \phi_{n-1}} + e^{-\beta \phi_{n}}}\nonumber \\
&&+ p_{n-1}(t)\lambda_{n-1}\frac{e^{-\beta \phi_{n}}}{e^{-\beta \phi_{n-1}} + e^{-\beta \phi_{n}}}
+p_{n+1}(t)\lambda_{n+1}'\frac{e^{-\beta \phi_{n}}}{e^{-\beta \phi_{n}} + e^{-\beta \phi_{n+1}}}
\end{eqnarray}
or, solely in terms of $\lambda_n$ using detailed balance ($\lambda_{n+1}' = \lambda_{n}$),
\begin{eqnarray}
\frac{\partial p_n(t)}{\partial t} &=& -p_n(t)\lambda_n\frac{e^{-\beta \phi_{n+1}}}{e^{-\beta \phi_{n}} + e^{-\beta \phi_{n+1}}}- p_n(t)\lambda_{n-1}\frac{e^{-\beta \phi_{n-1}}}{e^{-\beta \phi_{n-1}} + e^{-\beta \phi_{n}}}\nonumber \\
&&+ p_{n-1}(t)\lambda_{n-1}\frac{e^{-\beta \phi_{n}}}{e^{-\beta \phi_{n-1}} + e^{-\beta \phi_{n}}}
+p_{n+1}(t)\lambda_{n}\frac{e^{-\beta \phi_{n}}}{e^{-\beta \phi_{n}} + e^{-\beta \phi_{n+1}}}.\label{eq:discFPE}
\end{eqnarray}

We next introduce a lattice spacing $\epsilon$ and a continuous coordinate $x=n\epsilon$. We rewrite Eq.~\eqref{eq:discFPE} as
\begin{eqnarray}
\frac{\partial p(x,t)}{\partial t} &=& -p(x,t)\lambda(x)\frac{e^{-\beta \phi(x+\epsilon)}}{e^{-\beta \phi(x)} + e^{-\beta \phi(x+\epsilon)}}- p(x,t)\lambda(x-\epsilon)\frac{e^{-\beta \phi(x-\epsilon)}}{e^{-\beta \phi(x-\epsilon)} + e^{-\beta \phi(x)}}\nonumber \\
&&+ p(x-\epsilon,t)\lambda(x-\epsilon)\frac{e^{-\beta \phi(x)}}{e^{-\beta \phi(x-\epsilon)} + e^{-\beta \phi(x)}}
+p(x+\epsilon,t)\lambda(x)\frac{e^{-\beta \phi(x)}}{e^{-\beta \phi(x)} + e^{-\beta \phi(x+\epsilon)}}.
\end{eqnarray}
An expansion to order $\epsilon^2$ gives
\begin{equation}
\frac{\partial p(x,t)}{\partial t}= \frac{\epsilon^2}{2}\frac{\partial}{\partial x}\left[\lambda(x)\left(\beta \frac{d\phi}{dx}p(x,t)+\frac{\partial p(x,t)}{\partial x}\right)\right].
\end{equation}
Thus, we identify $\lambda(x)$ as 
\begin{equation}
\label{eq:rate_lambda}
\lambda(x)=\frac{2}{\epsilon^2}D(x),
\end{equation}
where $D(x)$ is the diffusion constant, and obtain exactly the FPE we wish to simulate, Eq.~\eqref{eq:FPE}.

\subsection{Simulation walkthrough}\label{sec:walkthrough}

\subsubsection{Initial conditions}\label{sec:IC}

We pick a large number of ``background'' particles $N$, so that we simulate $N+1$ particles in total that diffuse with free boundary conditions. Their positions are stored in an array. Following notation established earlier, denote the number of particles that are initially placed to the left of the interface (at $x=0$) as $N_\L$, and to the right as $N_\R=N-N_\L$. Choose a final running time $\tf$.

In the annealed case. the particles are  placed with a uniform distribution with mean densities $\brho_\L$ on the left and $\brho_\R$ on the left. Due to the free boundary conditions, the particles on the outermost edges may escape, and so the system will become diluted over time far from the interface. However we do not want  the tracer to be affected by this finite-size dilution effect. A particle beginning at the origin will typically go as far as $\sqrt{2D_\L t}$ to the left and $\sqrt{2D_\R t}$ to the right. Thus, so that it would be very unlikely for the tracer to reach the ever-diluting boundary, we chose $N_\L\gg\brho_\L\sqrt{2D_\L \tf}$ and $N_\R\gg\brho_\R\sqrt{2D_\R \tf}$.

We will perform both annealed ideal-gas and crystalline (mimicking quenched) simulations, so now we present how to create the initial configuration.

We begin with the ideal-gas case. Since we put two open systems in contact, the equilibrium condition requires to have the correct fluctuations in the number of particles. Namely, were we just to choose $N_\L=Ne^{-\beta\phi_\L}/(e^{-\beta\phi_\L}+e^{-\beta\phi_\R})$ and $N_\R=Ne^{-\beta\phi_\R}/(e^{-\beta\phi_\L}+e^{-\beta\phi_\R})$, this would describe two closed systems that were suddenly placed in contact\,---\,a nonequilibrium state. Our theory assumes pre-equilibration of the two bulks. Coupled with the very slow approach to our late-time asymptotics, this leads to an underestimation of the drift and mean-squared displacement. This has also manifested in the number of crossings being undersampled (especially the ones from high to low potential). 

Therefore, we choose $N=5\times(\brho_\L+\brho_\R)\sqrt{2\mathrm{max}(D_\L,D_\R)\tf}$. Then, starting from $N_\L=N_\R=0$, via a loop starting at 1 and ending at $N$, we change $N_\L\to N_\L+1$ with probability $e^{-\beta\phi_\L}/(e^{-\beta\phi_\L}+e^{-\beta\phi_\R})$ or $N_\R\to N_\R+1$ with the complementary probability $e^{-\beta\phi_\R}/(e^{-\beta\phi_\L}+e^{-\beta\phi_\R})$. Through this random assignment of $N_\L$ and $N_\R$, we mimic a case where we first waited for the two separate closed systems (with a total number of particles $N$) to equilibrate and thus have the right fluctuation statistics for the number of particles on either side (in a grand-canonical ensemble). Once this is done, for the purpose of a faster simulation, we `truncate' the number of particles $N_{j: D_j=\mathrm{min}(D_\L,D_\R)}$ in the bulk of the smaller diffusivity by a factor $\sqrt{\mathrm{min}(D_\L,D_\R)/\mathrm{max}(D_\L,D_\R)}$ (as the resulting number of particles will suffice to address the dilution problem). Lastly, within each bulk, we randomly place the $N_\L$ and $N_\R$ particles in a uniform distribution within $[-N_\L/\brho_\L,0)$ and $[0,N_\R/\brho_\R)$, and assign the remaining $N+1$th position at $0$ for the SF tracer. Their sorted positions in increasing order will be inserted in a positions array $\{X_n(t)\}$, such that $\{X_{1\leq n\leq N_\L}(0)\}<0$, $X_{N_\L+1}=0$, and $\{X_{N_\L+1\leq n\leq N_\L+N_\R+1}(0)\}\geq0$.

To simulate the quenched case, in theory, one must choose an initial configuration generated according to the above prescription, and perform many noise realizations over any initial configuration. This is not practical. Instead, as we saw in Sec.~\ref{sec:hyperuniform}, we will emulate the quenched statistics by performing an annealed simulation with crystalline initial conditions while suppressing any number-of-particles fluctuations. Namely, we simply choose $N_\L=5\times\brho_\L\sqrt{2D_\L\tf}$ and $N_\R=5\times\brho_\R\sqrt{2D_\R\tf}$. Then, we initially place the $N_\L+1+N_\R$ particles on the grid $-n/\brho_\L$ ($1\leq n\leq N_\L$), $0$, and $m/\brho_\R$ ($1\leq m\leq N_\R$), and save these initial positions within the array $\{X_n(t)\}$.

With the so prepared initial conditions of $N+1$ positions $\{X_n(0)\}$, we now propagate the particles as if they are diffusing independently.

\subsubsection{Parallel isolated-particle simulation}

First, we must pick a discretization $\epsilon$. Since this simulation scheme does not require the evaluation of derivatives of potentials and diffusivity, it is expected to work perfectly well for problems with potential and diffusivity discontinuities without needing to, \eg smooth out the potential~\cite{Farago20}. 
So to accurately sample external potentials and diffusivities, in theory, the lengthscales over which these quantities vary must be much larger than $\epsilon$. This does mean, however, that in the present problem of a single interface, owing to its self-similarity $x/t^{1/2}$ [see Eq.~\eqref{eq:sol_NORMAL} below], $\epsilon$ can be \emph{arbitrarily big} for large $\tf$\,---\,it should simply be much smaller than the typical length that the particles traverse by $\tf$, $\sqrt{2\mathrm{max}(D_\L,D_\R)\tf}$. In a multilayered systems~\cite{FP20}, again $\epsilon$ should be smaller that the distance among interfaces.
Thus, for the range of $\tf$ chosen below, with each $10$-fold increase in $\tf$, one may increase $\epsilon$ by $\sqrt{10}$ without affecting the accuracy at all. We will specify the values of $\epsilon$ later.

Upon choosing an $\epsilon$, with the given $D(x)$ and $\phi(x)$ [Eqs.~\eqref{eq:dif} and~\eqref{eq:pot}], one determines$\lambda(x)$ [Eq.~\eqref{eq:rate_lambda}] and $W(x\to x\pm\epsilon)$ [Eqs.~\eqref{eq:rate_W} and~\eqref{eq:rate_Wrev}, where $\lambda'(x+\epsilon)=\lambda(x)$]. Since the particles are not initially distributed on lattice (or, in the quenched case, not necessarily on the lattice set by $\epsilon$-spaced  jumps), while a normal-diffusing tracer will appear as if it jumps on a lattice, the SF tracer will move on a continuum.

Now, for each particle $n=1,\ldots,N+1$ independently, we perform the following. Define a timer variable $\tau_n$ which is initially set to $\tau_n=0$. Each single-particle simulation itself consists of a series of time-steps indexed by $k=1,2,\ldots$. (The total number of time-steps for each particles may be different.) By the end of the $k$th time-step, the particle is positioned at $x^k_n$ and its timer has reached $\tau_n=\tau^k_n$. Now, the $k+1$ time-step goes as follows: Given the particle's latest position, $x^k_n$, pick a waiting time $\Delta \tau^{k+1}_n$ from an exponential distribution with the rate parameter $W_\mathrm{tot}(x^k_n)=W(x^k_n\to x^k_n+\epsilon)+W(x^k_n\to x^k_n-\epsilon)$, the total jump rate. [This can be done using, \eg the inverse transform sampling method, $\Delta\tau^{k+1}_n=-[1/W_\mathrm{tot}(x^k_n)]\ln (1-u^{k+1}_n)$, where $u^{k+1}_n$ is uniformly distributed within $[0,1)$.] Then, its timer will be changed to $\tau_n=\tau^{k+1}_n=\tau^k_n+\Delta\tau^{k+1}_n$, and its position will be modified to either $x_n^{k+1}=x_n^k+\epsilon$ [with probability $W(x_n^k\to x_n^k+\epsilon)/W_\mathrm{tot}(x_n^k)$] or $x_n^{k+1}=x_n^k-\epsilon$ [with probability $W(x_n^k\to x_n^k-\epsilon)/W_\mathrm{tot}(x_n^k)$]. Once $\tau_n>\tf$, the simulation ends for this particle.

Suppose we wish to record the position of the $n$th particle. Then, its position at a time $t$ which is bounded by $\tau^k_n$ and $\tau_n^{k+1}$ will be taken  as $X_n(\tau_n^k\leq t<\tau_n^{k+1})=x_n^k$, up to errors of order $\epsilon$. 

\subsubsection{Data handling}

In the simulations below, we will report the system configuration at a few equidistant times between $0$ and $\tf$. First, to verify our method, we report the position $X_{N_\L+1}(t)$, \ie the isolated particle that started from $X_{N_\L+1}(0)=0$. This particle will diffuse normally (according to the results of Sec.~\ref{sec:normaldif} below). Next, we export the number of crossings to either side\,---\,$N_+$ by counting the number of $X_{1\leq n\leq N_\L}(t)\geq0$ and $N_-$ by the number of $X_{N_\L+1\leq n\leq N_\L+N_\R+1}(t)\leq0$. Lastly we also report the positions of the SF tracer. For this purpose, we sort the array $\{X_n(t)\}$ into a separate array, $\{Y_n(t)\}$ (rather than overwriting $\{X_n(t)\}$), and export the position of the $N_\L+1$th particle from the left, that is, $Y_{N_\L+1}(t)$.

\subsection{Normal diffusion of an isolated tracer}\label{sec:normaldif}

We briefly recall the solution to the diffusion of an isolated tracer starting at the origin~\cite{Farago20}. The probability distribution to find its position, $p_X(x,t|0,0)$, obeys the FPE [Eq.~\eqref{eq:FPE}] with initial conditions $p(x,0|0,0)=\delta(x)$. First, separate
\begin{equation}
    p_X(x,t|0,0)	=	\begin{cases}
p_{X,\L}(x,t|0,0), & x<0,\\
p_{X,\R}(x,t|0,0), & x>0,
\end{cases}
\end{equation}
Again it is  convenient to Laplace transform: $\tilde p_X(x,s|0,0)=\int_0^\infty dte^{-st}p_X(x,t|0,0)$ and obtain the equations
\begin{eqnarray}
    s\tilde p_{X,\L}(x,s|0,0)&=&D_\L\frac{\partial^2 \tilde p_{X,\L}(x,s|0,0)}{\partial x^2},\\
    s\tilde p_{X,\R}(x,s|0,0)&=&D_\R\frac{\partial^2 \tilde p_{X,\R}(x,s|0,0)}{\partial x^2},
\end{eqnarray}
while ensuring continuity, $e^{\beta\phi_\L}\tilde{p}_{X,\L}(0^{-},s|0,0)=e^{\beta\phi_\R}\tilde{p}_{X,\R}(0^{+},s|0,0)$ and $D_\R(\partial \tilde{p}_{X,\R}/\partial x)|_{0^{+},s}=1+D_\L(\partial \tilde{p}_{X,\L}/\partial x)|_{0^{-},s}$ (the $1$ comes from the delta initial conditions). Inverting the Laplace transform then gives
\begin{equation}
    p_X(x,t|0,0)=\begin{cases}
\dfrac{e^{-\beta\phi_\L}}{e^{-\beta\phi_\R}\sqrt{D_\R}+e^{-\beta\phi_\L}\sqrt{D_\L}}\dfrac{1}{\sqrt{\pi t}}\exp\left[-\dfrac{x^2}{4D_\L t}\right], & x<0,\\
\dfrac{e^{-\beta\phi_\R}}{e^{-\beta\phi_\R}\sqrt{D_\R}+e^{-\beta\phi_\L}\sqrt{D_\L}}\dfrac{1}{\sqrt{\pi t}}\exp\left[-\dfrac{x^2}{4D_\R t}\right], & x>0.
\end{cases}\label{eq:sol_NORMAL}
\end{equation}
From this, we find the drift and mean-squared displacement of an isolated tracer,
\begin{eqnarray}
    \langle X(t)\rangle&=&2\frac{D_\R/D_\L-e^{\beta(\phi_\R-\phi_\L)}}{(D_\R/D_\L)^{1/2}+e^{\beta(\phi_\R-\phi_\L)}}\sqrt{\frac{D_\L t}{\pi}},\label{eq:isoDRIFT}\\
    \langle X^2(t)\rangle_{\c}&=&2\left\{\frac{(D_\R/D_\L)^{3/2}+e^{\beta(\phi_\R-\phi_\L)}}{(D_\R/D_\L)^{1/2}+e^{\beta(\phi_\R-\phi_\L)}}-\frac{2}{\pi}\left[\frac{D_\R/D_\L-e^{\beta(\phi_\R-\phi_\L)}}{(D_\R/D_\L)^{1/2}+e^{\beta(\phi_\R-\phi_\L)}}\right]^2\right\}D_\L t.\label{eq:isoMSD}
\end{eqnarray}
These are Eqs.~\eqref{leteq:isoDRIFT} and~\eqref{leteq:isoMSD} of the main text.

\subsection{Results}\label{sec:results}

We show the simulation results for the normally-diffusing tracer and ideal-gas (annealed) and crystalline (quenched) number of crossings. The results concerning the ideal-gas and crystalline SF tracer were shown in Figs.~\ref{fig:sfd}--\ref{fig:sfdTRANS} of the main text. For the graphs below, we pick the parameters $D_\L=1$, $\beta\phi_\L=0$, and $\brho_\L=2.515$ for the left bulk, and $D_\R=3$, $\beta\phi_\R=3$, and $\brho_\R=0.125$ for the right one, and made $2\cdot10^4$ samples. 

We perform a set of simulations of varying final times, $\tf\in\{10^{-2},10^{-1},1,10,100,10^3,10^4,10^5,10^6,10^7,10^8\}$. During each simulation, we export the above data at $20$, equispaced times. We separated these simulations as we may accelerate the longer ones without loosing precision by choosing the following, increasing lattice spacings, respectively to the finl times $\tf$: $\epsilon\in\{10^{-3},10^{-3},10^{-2},10^{-2},10^{-1},10^{-1},1,1,10,10,100\}$. (Notice the incredibly course discretization for $\tf=10^8$\,---\,$\epsilon=100$.) As a result of the increasing $\tf$, the total number of particles also increases as $\sqrt{10}$ between each $\tf$ (so to address the dilution problem, see Sec.~\ref{sec:IC}).

\begin{figure}
    \centering
    \includegraphics[width=0.49\linewidth]{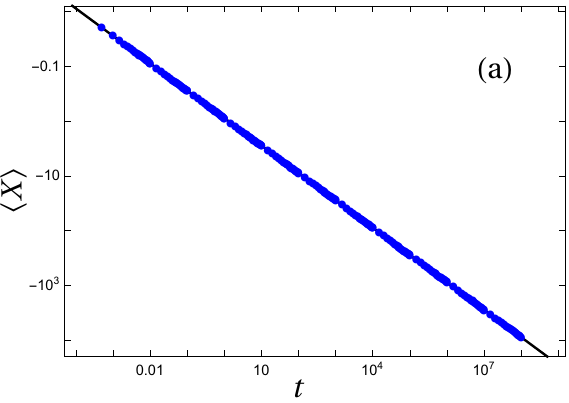}
    \includegraphics[width=0.49\linewidth]{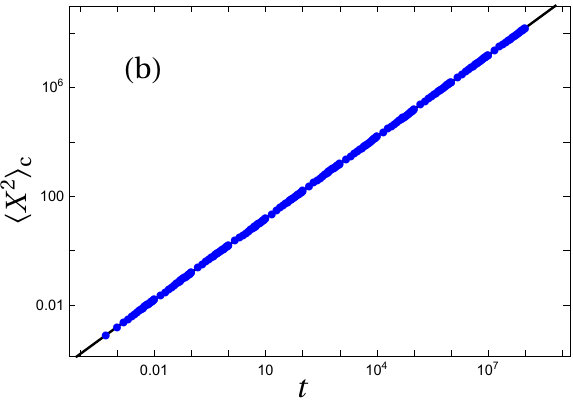}
    \caption{Simulation. The (a) drift $\langle X\rangle$ and (b) mean-squared displacement $\langle X^2\rangle_\c$ of an isolated (and hence) normal-diffusing tracer particle whose initial position was the interface, $X(0)=0$. The blue points are given by the simulations, while the black lines are the theoretical predictions [Eq.~\eqref{eq:isoDRIFT} and~\eqref{eq:isoMSD}, respectively]. Parameter values: $D_\L=1$, $\beta\phi_\L=0$, $\brho_\L=2.515$, and $D_\R=3$, $\beta\phi_\R=3$, and $\brho_\R=0.125$, as in the rest of the figures.}
    \label{fig:norm}
\end{figure}

\begin{figure*}
    \centering
    \includegraphics[width=0.49\linewidth]{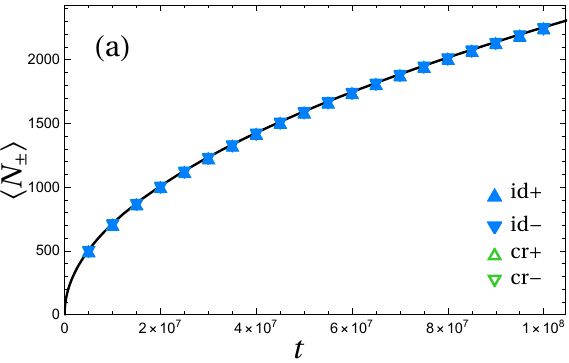}
    \includegraphics[width=0.49\linewidth]{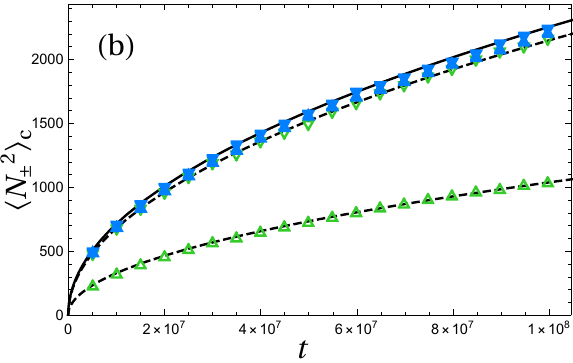}
    \caption{Simulation. Ideal-gas (`$\id$') and crystalline (`$\crr$') (a) mean $\langle N_\pm\rangle$ and (b) second cumulant $\langle N^2_\pm\rangle_\c$ of the number of crossings right- (`$+$') and leftwards (`$-$') at late times. The simulation results are depicted by the labels indicated in panel (a), while the immediately adjacent lines are the theoretical predictions [Eqs.~\eqref{eq:NavA} and~\eqref{eq:NavQ}]. Parameter values are as in Fig.~\ref{fig:norm}.}
    \label{fig:cross}
\end{figure*}

Figure~\ref{fig:norm} shows the results for the above mentioned separate simulations for the isolated tracer that was initially positioned at the interface, $X(0)=0$. In the notation of Sec.~\ref{sec:walkthrough}, we show the  drift [Fig.~\ref{fig:norm}(a)] and mean-squared displacement [Fig.~\ref{fig:norm}(b)] of the $N_\L+1$th entry in the unsorted array of positions $\{X_n(t)\}$. Indeed the simulation is very accurate despite the large $\epsilon$ at later times, owing to the self-similarity of the problem. In the context of Fig.~\ref{fig:sfdTRANS} of the main text, notice that the particle drift is leftwards for these parameters, in contrast to the rightwards (uphill) drift reported for the SF tracer.

Figure~\ref{fig:cross} shows the ideal-gas and crystalline means [Fig.~\ref{fig:cross}(a)] and second cumulants [Fig.~\ref{fig:cross}(b)] of the number of crossings. Once again, the simulation produces the anticipated results with a high precision even with $\epsilon=100$ used for the presented times. With that, we verified the validity and accuracy of the proposed simulation method. The results for the SF tracer are presented in the main text in Figs.~\ref{fig:sfd}--\ref{fig:sfdTRANS}.

\end{widetext}


\begin{thebibliography}{999}
\bibitem{FOOTmft} This is justified at late times by the rescaling into a uniform constant diffusivity. For details, see the treatments of SFD based on, \eg the macroscopic fluctuation theory~\cite{KMS14,BDGJL01,BDGJL05,BDGJ06} or an homogenization argument~\cite{SD23,BS10}.
\bibitem{SM} See {S}upplementary {M}aterial for the full derivation of the main result and for further details on the simulation. It includes Refs.~\cite{sal18,OKS03,FP20}.
\bibitem{FOOTid} This is justified due to the zero range of the hard-core repulsion.
\end{thebibliography}
\end{document}